\documentclass[final,5p,times,twocolumn]{elsarticle}

\usepackage[utf8]{inputenc}
\usepackage[T1]{fontenc}
\usepackage[english]{babel}
\usepackage{graphicx}
\usepackage{amsmath}
\usepackage{amsfonts}
\usepackage{amssymb}
\usepackage{booktabs}
\usepackage{tabularx}
\usepackage{microtype}
\usepackage{placeins}
\usepackage{xcolor}
\usepackage[colorlinks=true, allcolors=blue]{hyperref}
\usepackage{filecontents}
\usepackage{caption}
\usepackage{stfloats}

\begin{document}

\begin{frontmatter}

\title{Phonon-mediated visible upconversion luminescence in femtosecond 1030-nm pumped $\text{Tm}^{3+}$-doped silica fibers}

\author[aff1]{Nicol\`o Pisani\corref{cor1}}
\ead{nicolo.pisani@unical.it}
\author[aff1]{Raffaele Filosa}
\author[aff1]{Alessia Cirimele}
\author[aff1]{Lucas Zanaglia}
\author[aff2]{Jan Aubrecht}
\author[aff2]{Michal Kamr\'adek}
\author[aff2]{Ivan Ka{\v{s}}{\'i}k}
\author[aff2]{Ivo Barto\v{n}}
\author[aff2]{Ond\v{r}ej Podrazk\'y}
\author[aff2]{Jan Mr\'azek}
\author[aff1]{Mario Ferraro\corref{cor1}}

\cortext[cor1]{Corresponding authors}

\affiliation[aff1]{organization={Dipartimento di Fisica},
  addressline={Universit\`a della Calabria, Ponte Pietro Bucci Cubo 33 C}, 
  city={Rende (CS)},
  postcode={87036}, 
  country={Italy}}

\affiliation[aff2]{organization={Institute of Photonics and Electronics of the Czech Academy of Sciences},
  addressline={Chabersk\'a 57}, 
  city={Prague 8},
  postcode={182 51}, 
  country={Czech Republic}}

\begin{abstract}
We investigate multiphoton absorption-driven visible luminescence in a thulium-doped aluminosilicate optical fiber excited by an ultrafast 1030-nm femtosecond pulse train in the {MHz domain}. We report the observation of an unusual spectral crossover: the predominance of an emission band centered at 457-nm (deep blue) is progressively overtaken by a second emission, centered at 475-nm (cyan), as launched average power increases. Concurrently, a third near-infrared emission around 792~nm exhibits persistent, monotonic growth characterized by a well-defined bilinear slope transition. By combining calibrated spectral band integration with a structure-informed rate-equation model based on host thermodynamics, we show that the visible spectra evolution is governed by non-radiative multi-phonon transfer and phonon-assisted cross-relaxation activated by cumulative core heating, rather than permanent color-center solarization. These results offer new predictive insights into the thermal and spectroscopic stability of high-power ultrafast fiber systems.
\end{abstract}


\end{frontmatter}

\section{Introduction}

Thulium-doped silica fibers have emerged as a versatile platform for a wide range of applications, spanning from mid-infrared laser sources for medical surgery and gas sensing to high-power ultrafast amplifiers \cite{digo2001, khamis2016, duclos1995, varak2024, gaida2017}. $\text{Tm}^{3+}$-doped fibers are attractive because of their broad emission band around 2~$\mu\text{m}$ and the efficiency of the so-called ``two-for-one'' cross-relaxation (CR) process, which allows for quantum efficiencies exceeding unity under 790-nm pumping \cite{yinzi2020, tao2013, moulton2009, Paschotta__quantum_efficiency}.

Thulium-doped materials also display significant upconversion luminescence (UL) in the visible and ultraviolet ranges, driven by multiphoton absorption or energy transfer processes \cite{simpson2008, tanabe1998, auzel2004}. 
Although the development of blue fiber lasers has been a historical motivation for studying these visible emissions \cite{barber1994, duclos1995, zhao2010}, the visible spectrum is increasingly valued as a diagnostic tool to characterize population dynamics, ion clustering, and material degradation in high-power systems \cite{frith2010, pal2013, jlumin2023}.

Blue upconversion in thulium-doped silica has been reported under various excitation regimes, most notably using 980-nm or 1064-nm pumping \cite{simpson2008, yang2018, xia2015}. The blue emission generally involves two main features: a band in the 450--460~nm range, related to the ${}^1\mathrm{D}_2 \rightarrow {}^3\mathrm{F}_4$ electronic transition, and a band in the 470--480~nm range, related to the ${}^1\mathrm{G}_4 \rightarrow {}^3\mathrm{H}_6$ transition \cite{tanabe1998, buckthorpe2025, kamradek2021, gaida2017, unger2025}. The population of these levels requires the sequential absorption of three to five photons, subject to the pump wavelength and the specific intermediate levels involved. For instance, a 980-nm excitation implies a three-step process to populate the ${}^1\mathrm{G}_4$ level \cite{simpson2008}. In contrast, exciting the system with {IR wavelengths} involves a different set of resonances, often requiring phonon-assisted processes to bridge the energy gaps between the thulium levels and the pump wavelength \cite{kamradek2021, messias2002}.

A fundamental but seldom studied aspect of these mechanisms is the relative evolution of the blue emission peaks under high-intensity, ultra-short pulse excitation. Previous studies by Tanabe et al. \cite{tanabe1998} under continuous-wave pumping at 630--670~nm in high-germania silica fibers reported that the relative branching ratio between the 455-nm and 480-nm bands varied with launched power. This redistribution was interpreted as an athermal optical stepping effect driven by differential ground-state depletion and excited-state absorption (ESA) saturation within the $\text{GeO}_2\text{--}\text{SiO}_2$ host \cite{tanabe1998}. In germanosilicate fibers, the presence of substantial germanium concentrations ($\ge 15\text{--}25\,\text{mol}\%$) introduces dense distributions of germanium-related oxygen-deficient centers (GODCs), accelerating photochemical defect activation and providing athermal virtual pathways, as observed under ultrafast pulses \cite{ferraro2021prj}. Beyond conventional rare-earth-mediated energy-transfer mechanisms, high-peak-power femtosecond pulses in the IR domain can directly induce high-order multiphoton absorption in silica and soft glass optical fibers, promoting the population of high-lying electronic states and defect centers \cite{mangini2020, ferraro2024}. Building on this mechanism, studies have shown that intense femtosecond pulse propagation in silica-based fibers can occur in a deterministic sub-breakdown regime, where nonlinear absorption governs optical power transmission and drives intense defect-mediated visible photoluminescence without causing irreversible material damage \cite{ferraro2021prj}. Complementarily, spatially resolved investigations of ultrashort-field propagation in guided silica geometries have enabled the mapping of transverse field dynamics and the tracking of the associated luminescent defects, revealing the sensitivity of visible emission to local field confinement \cite{ferraro2025}.

However, when transitioning from germanosilicate matrices to heavily doped aluminosilicate optical fibers, the structural and thermodynamic behavior changes fundamentally. In aluminosilicate matrices prepared via modified chemical vapor deposition, $\text{Al}_2\text{O}_3$ polyhedra directly encapsulate rare-earth cations, shifting the effective local phonon spectrum and significantly suppressing native point defects \cite{kamradek2019jphot, leich2022oe, peterka2011}. Concurrently, when excited by megahertz femtosecond pulse trains ($f_{\text{rep}} \sim 1\,\text{MHz}$), the microsecond pulse separation ($t_{\text{inter}} \approx 1\,\mu\text{s}$) becomes much shorter than the core thermal diffusion time ($\tau_{\text{diff}} \sim 12\,\mu\text{s}$) \cite{eaton2005}. The aluminosilicate core enters a cumulative thermal regime, raising the steady-state core temperature by several tens to hundreds of Kelvin \cite{eaton2005, lancry2013}. While visible luminescence has occasionally been linked to photodarkening via multi-photon color-center generation \cite{frith2010, elagmy2014, broer1993, broer1993ol}, {operating below catastrophic dielectric breakdown ensures thermal multi-phonon emission decisively supersedes irreversible photochemical solarization}. Furthermore, the high repetition rate makes energy transfer a non-stationary process where pulse separation becomes comparable to intermediate state lifetimes \cite{messias2002, lupi2014}.

To address these coupled phenomena, researchers have traditionally relied on empirical systems of rate equations constructed around phenomenological cross-relaxation and energy transfer parameters \cite{fu2010, tse2015, simpson2006, jackson1998, pal2010, lupi2016, kamradek2021}. Recent work by Buckthorpe and Jackson has underlined the need to refine these models for aluminosilicate glass hosts \cite{buckthorpe2025, barber1994}. Incorporating alumina via modified chemical vapor deposition introduces localized $[\text{AlO}_4]^-$ coordination polyhedra that modify the effective phonon cutoff energy and alter non-radiative lifetimes \cite{kamradek2019jphot, leich2022oe, peterka2011}. Neglecting the interplay between thermal phonon accumulation and the altered local coordination shell in aluminosilicate fibers prevents conventional models from reproducing power-dependent crossovers and reversible visible roll-offs.

In this work, we investigate multiphoton visible upconversion and near-infrared dynamics in a heavily $\text{Tm}^{3+}$-doped aluminosilicate fiber excited by ultrafast laser pulses at 1030~nm. Through a set of experimental checks---including bidirectional power cycling, repetition-rate scaling, and spectral band integration---we demonstrate that the reversible visible roll-off and the observed crossover between the 457-nm (${}^1\mathrm{D}_2$) and 475-nm (${}^1\mathrm{G}_4$) lines serve as direct signatures of thermal population redistribution. By combining these measurements with a structure-informed rate-equation model based on host thermodynamics and nonlinear guided-wave optics, we elucidate the interplay between multi-photon pumping, cumulative heat accumulation, and phonon-assisted cross-relaxation.

\begin{figure*}[!t]
  \centering
  \includegraphics[width=0.95\textwidth]{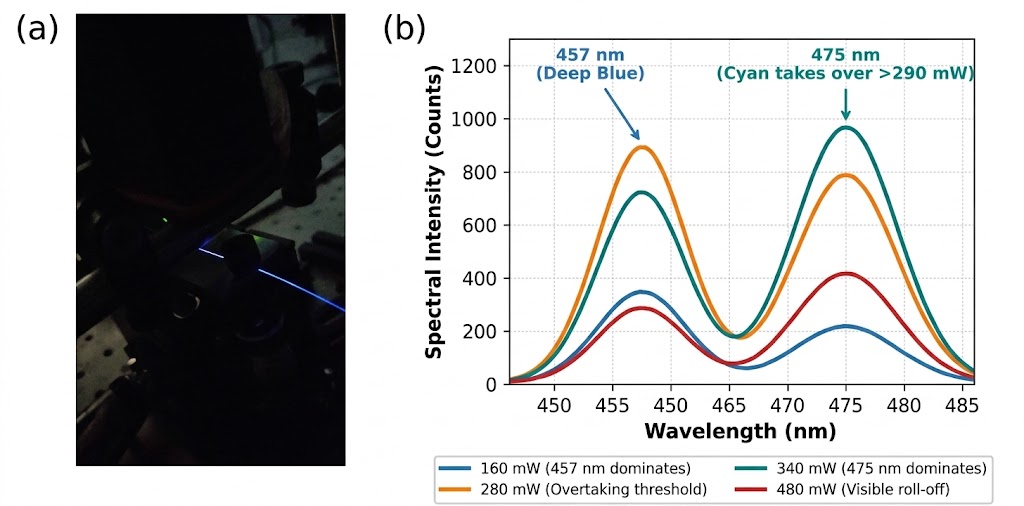}
  \caption{Experimental observation of guided upconversion luminescence: (a) Photograph of the \texorpdfstring{$\text{Tm}^{3+}$}{Tm3+}-doped aluminosilicate fiber intensely glowing blue along the propagation length under ultrafast 1030-nm femtosecond excitation in the dark; (b) Power-dependent visible emission spectra highlighting the competitive population branching between the Deep Blue \texorpdfstring{(${}^1\mathrm{D}_2 \rightarrow {}^3\mathrm{F}_4$, 457~nm)}{(1D2 to 3F4, 457 nm)} and Cyan \texorpdfstring{(${}^1\mathrm{G}_4 \rightarrow {}^3\mathrm{H}_6$, 475~nm)}{(1G4 to 3H6, 475 nm)} bands, documenting the spectral crossover threshold around \texorpdfstring{$\sim 283\text{--}290\,\text{mW}$}{~283--290 mW} and the subsequent thermal roll-off.}
  \label{fig:glow}
\end{figure*}

Figure~\ref{fig:glow} displays the active fiber glowing bright blue in the dark during femtosecond pumping. The visual luminescence exhibits the spatial uniformity of the guided emission along the fiber length, while the plot illustrates the underlying population branching between the high-lying ${}^1\mathrm{D}_2$ level (457~nm) and the ${}^1\mathrm{G}_4$ level (475~nm) that governs the visible spectrum.

\section{Materials and methods}

\subsection{Fiber structural and compositional profilometry}
The active optical fiber was fabricated using MCVD coupled with the nanoparticle-doping technique \cite{kamradek2019jphot}. Porous silica soot core layers were soaked with an ethanolic suspension of $\gamma$-$\text{Al}_2\text{O}_3$ nanoparticles ($<50\,\text{nm}$) and $\text{TmCl}_3$, followed by chlorine gas vitrification and preform collapse above $1950\,^\circ\text{C}$ \cite{kamradek2019jphot}. During vitrification, alumina nanoparticles dissolve into the vitreous silica network, forming an amorphous aluminosilicate environment that ensures rare-earth ion dispersion and minimizes pair-induced quenching \cite{kamradek2019jphot, arai1986}. Background attenuation was measured between $10\text{--}20\,\text{dB/km}$ around 880-nm.

\begin{figure*}[!t]
  \centering
  \includegraphics[width=0.98\textwidth]{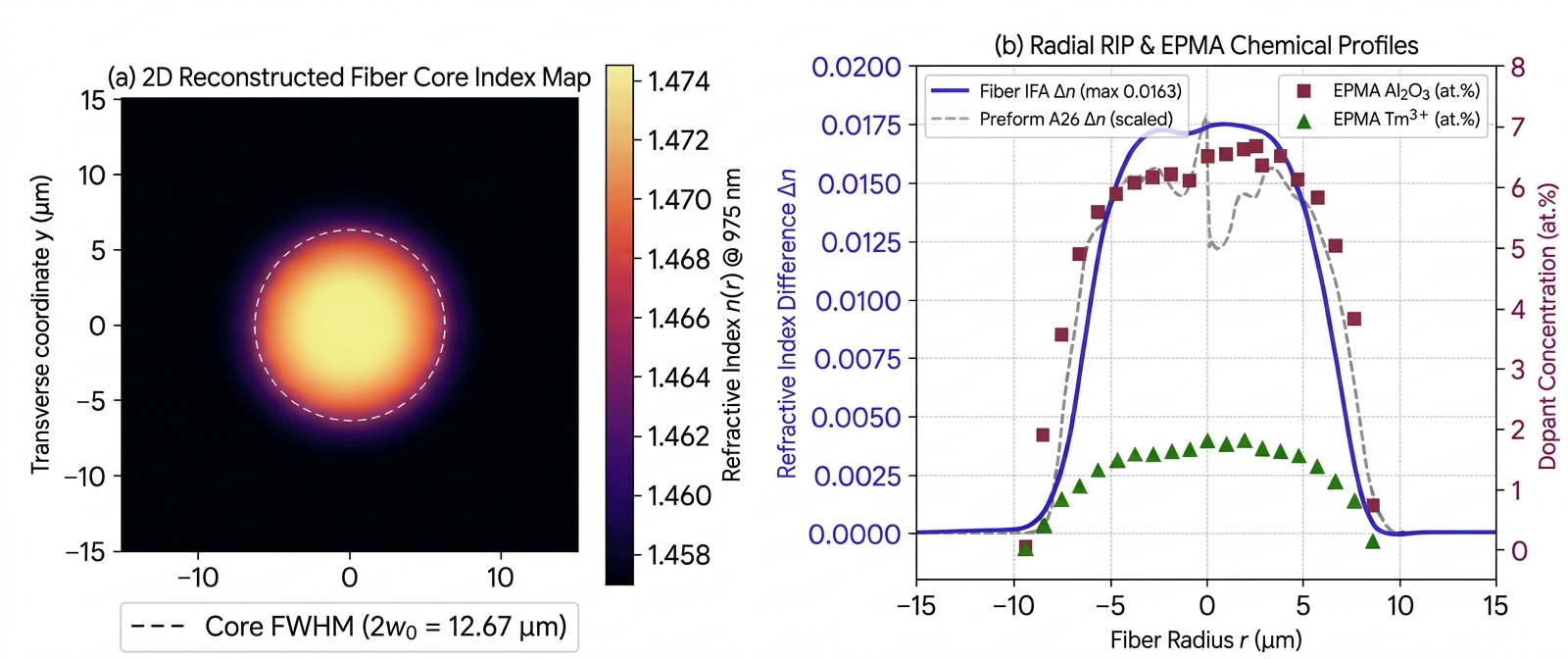}
  \caption{Structural and compositional profilometry of the thulium-doped fiber: (a) Reconstructed 2D cross-sectional refractive index map $n(x,y)$ of the core at $975\,\text{nm}$ ($n_{\text{clad}} = 1.4570$, $\text{NA} \approx 0.22$), with the dashed white line indicating the mode-field boundary ($2w_0 = 12.67\,\mu\text{m}$); (b) 1D radial refractive index profile $\Delta n(r)$ measured via IFA-100 profilometry and scaled preform profilometry, superimposed on Electron Probe Microanalysis (EPMA) atomic concentration profiles of $\text{Al}_2\text{O}_3$ (red squares) and $\text{Tm}^{3+}$ (green triangles).}
  \label{fig:index_profile}
\end{figure*}

Refractive index profilometry measured on an IFA-100 profiler (calibrated against pure silica cladding $n_{\text{clad}} = 1.4570$ at $975\,\text{nm}$) is shown in Fig.~\ref{fig:index_profile}(a) and (b). The core exhibits a maximum index step $\Delta n_{\text{max}} = 0.01634 \approx 0.017$, corresponding to a high core numerical aperture, $\text{NA} = \sqrt{(n_{\text{clad}} + \Delta n_{\text{max}})^2 - n_{\text{clad}}^2} \approx 0.22$.
The core cross-section displays a baseline diameter $d_{\text{max}} = 18.63\,\mu\text{m}$ and a central Gaussian-like full-width at half-maximum $\text{FWHM} = 12.67\,\mu\text{m}$. For excitation at $\lambda = 1030\,\text{nm}$, the effective mode radius ($w_0 = \text{FWHM}/2 \approx 6.335\,\mu\text{m}$) yields an effective modal area $A_{\text{eff}} = \pi w_0^2 \approx 1.261 \times 10^{-10}\,\text{m}^2$ and a normalized frequency parameter:
$V = \frac{2\pi}{\lambda} w_0 \text{NA} \approx 8.46$, confirming tight transverse mode confinement. Electron Probe Microanalysis (EPMA, Fig.~\ref{fig:index_profile}(b)) shows peak concentrations of $6.6\,\text{at.}\%$ $\text{Al}_2\text{O}_3$ and $1.82\,\text{at.}\%$ $\text{Tm}^{3+}$, corresponding to mode-averaged values of $4.1\,\text{at.}\%$ $\text{Al}_2\text{O}_3$ and $13872\,\text{ppm}$ atomic $\text{Tm}^{3+}$ ($N_{\text{Tm}} = 3.05 \times 10^{20}\,\text{ions/cm}^3$, atomic ratio $\text{Al/Tm} \approx 3.0$).

\subsection{Ultrafast laser excitation and guided field regime}
The excitation source was a commercial ultrafast Yb-doped femtosecond fiber laser delivering pulses centered at $\lambda = 1030\,\text{nm}$ with unchirped duration $\tau_{\text{FWHM}} = 173\,\text{fs}$, optical bandwidth $\Delta \lambda \approx 20\,\text{nm}$, and repetition rate $f_{\text{rep}} = 1008\,\text{kHz}$. The beam was coupled into a clean-cleaved fiber segment ($L \le 5.0\,\text{cm}$). Fiber, laser, and peak guided parameters are summarized in Table~\ref{tab:fiber_laser_params}.

\begin{table}[htbp]
\centering
\small
\caption{Structural parameters of the fiber, excitation laser parameters, and characteristic guided field values.}
\label{tab:fiber_laser_params}
\begin{tabular*}{\columnwidth}{@{\extracolsep{\fill}}lll@{}}
\toprule
\textbf{Parameter} & \textbf{Value} & \textbf{Unit} \\ 
\midrule
Cladding diameter ($d_{\text{clad}}$)                    & $128.0$                 & $\mu\text{m}$ \\
Cladding refractive index ($n_{\text{clad}}$)            & $1.4570$                & -- \\
Core baseline diameter ($d_{\text{max}}$)                 & $18.63$                 & $\mu\text{m}$ \\
Core mode diameter ($2w_0$)                               & $12.67$                 & $\mu\text{m}$ \\
Core refractive index step ($\Delta n_{\text{max}}$)     & $0.01634$               & -- \\
Numerical aperture ($\text{NA}$)                         & $0.22$                  & -- \\
Effective modal area ($A_{\text{eff}}$)                  & $1.261 \times 10^{-10}$ & $\text{m}^2$ \\
$\text{Tm}^{3+}$ mode-averaged density ($N_{\text{Tm}}$)  & $3.05 \times 10^{20}$   & $\text{cm}^{-3}$ \\
$\text{Al}_2\text{O}_3$ mode-averaged concentration      & $4.1$                   & $\text{at.}\%$ \\
Background attenuation at $1300\,\text{nm}$              & $10\text{--}40$         & $\text{dB/km}$ \\
Pump center wavelength ($\lambda$)                       & $1030$                  & $\text{nm}$ \\
Pump spectral bandwidth ($\Delta \lambda$)               & $20$                    & $\text{nm}$ \\
Unchirped pulse duration ($\tau_{\text{FWHM}}$)          & $173$                   & $\text{fs}$ \\
Laser repetition rate ($f_{\text{rep}}$)                 & $1008$                  & $\text{kHz}$ \\
Fiber interaction length ($L$)                           & $\le 5.0$               & $\text{cm}$ \\
Core launch coupling efficiency ($\eta_{\text{coup}}$)    & $0.55$                  & -- \\
Nonlinear Kerr index ($n_{2,\text{host}}$)               & $3.8 \times 10^{-20}$   & $\text{m}^2/\text{W}$ \\
Peak core power ($P_{\text{launch}} = 300\,\text{mW}$)     & $0.83$       & $\text{MW}$ \\
Peak core intensity ($P_\text{launch} = 300\,\text{mW}$) & $0.66$             & $\text{TW/cm}^2$ \\
\bottomrule
\end{tabular*}
\end{table}

For a characteristic launched average power $P_{\text{launch}} = 300\,\text{mW}$, the core power is $P_{\text{core}} = \eta_{\text{coup}} P_{\text{launch}} \approx 165\,\text{mW}$, yielding a single-pulse energy inside the core $E_{p,\text{core}} \approx 1.64 \times 10^{-7}\,\text{J}$. For a $\operatorname{sech}^2$ temporal pulse envelope, the peak power evaluates to $P_{\text{peak}} \approx 0.83\,\text{MW}$ and the peak on-axis intensity is $I_{\text{peak}} \approx 0.66\,\text{TW/cm}^2$. Across the full experimental pump range ($P_{\text{launch}} \le 480\,\text{mW}$), peak intensity remains strictly bounded below $I_{\text{peak}} \le 1.06\,\text{TW/cm}^2$. This hints that pulse propagation operates exclusively in the sub-breakdown multi-photon absorption regime of silica fibers \cite{ferraro2021prj}, well below catastrophic damage thresholds ($> 2\text{--}3\,\text{TW/cm}^2$), ensuring reversible spectroscopic interactions.

\subsection{Nonlinear multi-photon excitation and spectral broadening}
\begin{figure}[htbp]
  \centering
  \includegraphics[width=\columnwidth]{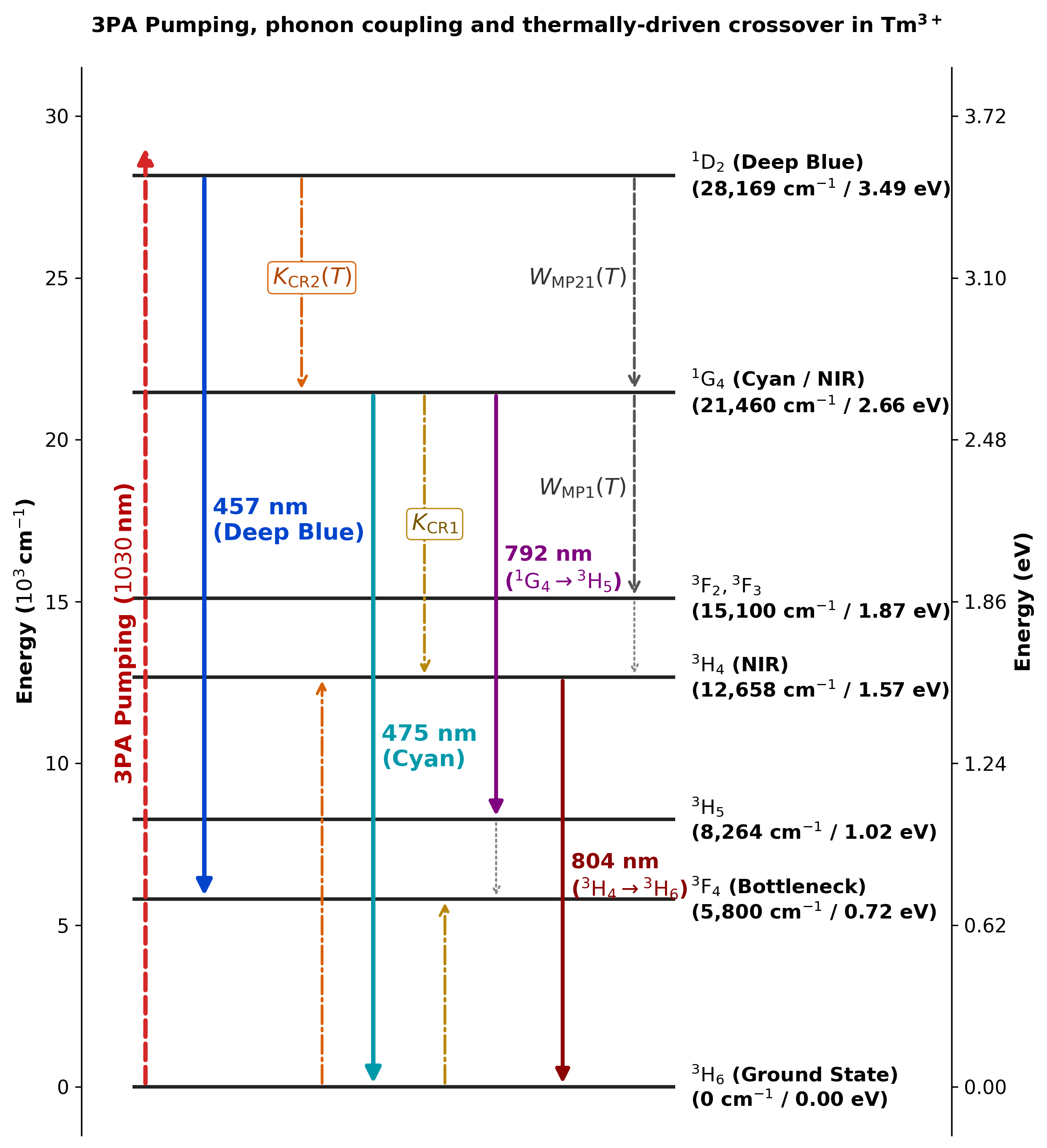}
  \caption{Energy-level diagram of $\text{Tm}^{3+}$ in aluminosilicate glass showing 1030-nm multi-photon pumping, visible emissions (457 and 475~nm), and the two distinct NIR pathways along with multi-phonon relaxation and phonon-assisted cross-relaxation. In the rate-equation model, both NIR transitions are resolved microscopically and integrated within the experimental 787--797~nm detection window via Eq.~(8).}
  \label{fig:levels}
\end{figure}

The electronic energy levels of $\text{Tm}^{3+}$ in aluminosilicate glass \cite{unger2025} are depicted in Fig.~\ref{fig:levels}. Under intense sub-picosecond irradiation, the absence of resonant real intermediate states at 1-photon and 2-photon energy intervals suppresses sequential excited-state absorption. Excitation of the upper visible manifold occurs via non-resonant multi-photon absorption directly intercepting virtual electronic states governed by nonlinear optical susceptibility \cite{boyd2020, he2008, mangini2020}.

The vibronic absorption manifold of ${}^1\mathrm{D}_2$ in aluminosilicate glass centers at $\lambda_{\text{abs}} = 355.0\,\text{nm}$ ($\tilde{\nu}_{\text{UV}} = 28169\,\text{cm}^{-1}$) \cite{unger2025}. {As the high-intensity pulse propagates along the core, Self-Phase Modulation (SPM) spectrally broadens the fundamental 1030-nm pump envelope, effectively generating spectral components at $3 \times 355.0\,\text{nm} = 1065.0\,\text{nm}$ that bridge into the three-photon absorption band of the ${}^1\mathrm{D}_2$ level.

Experimentally, double-logarithmic regression of the deep blue emission at low powers yields an effective power scaling of $m = 3.37 \pm 0.22$ ($R^2 = 0.983$, Fig.~\ref{fig:loglog_fit}). This slight deviation from pure cubic scaling is consistent with multi-photon absorption taking place under dynamic pulse broadening and Kerr mode confinement. In the macroscopic rate-equation framework, the pumping rate populating ${}^1\mathrm{D}_2$ is thus formulated directly as:
\begin{equation}
  R_{\text{MPA}} = {K_{\mathrm{MPA}} P_{\text{launch}}^{3.3}}
\end{equation}
where $K_{\mathrm{MPA}} = 4.71 \times 10^{-6}\,\text{mW}^{-3.3}\,\text{s}^{-1}$. As derived in \ref{app:k3pa}, this calibrated value accounts for the microscopic absorption cross-section, the SPM-broadened overlap at $1065\,\text{nm}$, and the experimental coupling efficiency.

\subsection{Rate-equation formulation}
Applying the quasi-steady-state approximation (see {\ref{app:qssa}}) to fast intermediate levels ${}^3\mathrm{F}_{2,3}$ and ${}^3\mathrm{H}_5$, the fractional populations $n_2$ (${}^1\mathrm{D}_2$), $n_1$ (${}^1\mathrm{G}_4$), and $n_4$ (${}^3\mathrm{H}_4$) obey the coupled system:
\begin{align}
  \frac{dn_2}{dt} &= R_{\text{MPA}} n_0 - \left(\gamma_{D2} + W_{\text{MP2}}\right) n_2 - K_{\text{CR2}} n_2 n_0 \label{eq:ode1} \\
  \frac{dn_1}{dt} &= \left(W_{\text{MP21}} + K_{\text{CR2}} n_0\right) n_2 \nonumber \\
                  &\quad - \left(\gamma_{G4,\text{cyan}} + \gamma_{G4,\text{nir}} + W_{\text{MP1}}\right) n_1 - K_{\text{CR1}} n_1 n_0 \label{eq:ode2} \\
  \frac{dn_4}{dt} &= K_{\text{CR2}} n_2 n_0 + \left(W_{\text{MP1}} + K_{\text{CR1}} n_0\right) n_1 - \gamma_{3H4} n_4 \label{eq:ode3}
\end{align}
The meaning and values of the parameters in the equations above are summarized in Table~\ref{tab:parameters}. Bottlenecking at ${}^3\mathrm{F}_4$ gives $n_3 \approx n_2 \gamma_{D2} / \gamma_{3F4}$, and ground-state closure is enforced via $n_0 = \max(0, 1 - n_1 - n_2 - n_3 - n_4)$. The collected 792-nm emission intensity is evaluated as:
\begin{equation}
  I_{792} \propto n_1 \gamma_{G4,\text{nir}} + \eta_4 n_4 \gamma_{3H4}
\end{equation}
with $\eta_4 = 0.28$ representing the spectral overlap of the $^3\mathrm{H}_4 \rightarrow {}^3\mathrm{H}_6$ band in the 787--797~nm window \cite{unger2025}. Non-radiative multi-phonon transfer rates $W_{\text{MP}}$ and cross-relaxation $K_{\text{CR2}}$ are coupled to core temperature via the Bose--Einstein thermal phonon occupancy factor. ({see \ref{app:thermal}} for greater details). 

\begin{table*}[!t]
\centering
\small
\caption{Physical parameters, rate coefficients, and physical ranges used in the rate-equation model.}
\label{tab:parameters}
\begin{tabularx}{\textwidth}{@{}l X c c X@{}}
\toprule
\textbf{Symbol} & \textbf{Physical Description} & \textbf{Value} & \textbf{Physical Range} & \textbf{Source / Justification} \\ 
\midrule
$\gamma_{D2}$              & Radiative rate ${}^1\mathrm{D}_2 \rightarrow {}^3\mathrm{F}_4$ ($457\,\text{nm}$) & $20000.0\,\text{s}^{-1}$               & $[18000, 22000]\,\text{s}^{-1}$ & Judd-Ofelt analysis ($\tau_{\text{rad}} \approx 50\,\mu\text{s}$) \cite{judd1962, ofelt1962, caird1975, unger2025} \\
$\gamma_{G4,\text{cyan}}$  & Radiative rate ${}^1\mathrm{G}_4 \rightarrow {}^3\mathrm{H}_6$ ($475\,\text{nm}$) & $3200.0\,\text{s}^{-1}$                & $[2800, 3600]\,\text{s}^{-1}$   & Cyan emission band \cite{unger2025} \\
$\gamma_{G4,\text{nir}}$   & Radiative rate ${}^1\mathrm{G}_4 \rightarrow {}^3\mathrm{H}_5$ ($792\,\text{nm}$) & $13000.0\,\text{s}^{-1}$               & $[11000, 15000]\,\text{s}^{-1}$ & Intermediate NIR branching ratio \cite{unger2025} \\
$\gamma_{3H4}$             & Radiative rate ${}^3\mathrm{H}_4 \rightarrow {}^3\mathrm{H}_6$ ($804\,\text{nm}$) & $1800.0\,\text{s}^{-1}$                & $[1500, 2200]\,\text{s}^{-1}$   & Intrinsic radiative rate \cite{caird1975, unger2025, peterka2011} \\
$\gamma_{3F4}$             & Bottleneck rate ${}^3\mathrm{F}_4 \rightarrow {}^3\mathrm{H}_6$                    & $2000.0\,\text{s}^{-1}$                & $[1800, 2500]\,\text{s}^{-1}$   & Terminal laser level ($\tau \approx 500\,\mu\text{s}$) \cite{jackson1998, unger2025} \\
$W_{0,2}$                 & Base multi-phonon rate ${}^1\mathrm{D}_2 \rightarrow {}^1\mathrm{G}_4$             & $23180.0\,\text{s}^{-1}$               & $[15000, 35000]\,\text{s}^{-1}$ & Energy gap law ($p = 7.0$), $[\text{AlO}_4]^-$ disorder \cite{layne1977, vandijk1983, leich2022oe, kamradek2019jphot} \\
$W_{0,1}$         & Base multi-phonon drain ${}^1\mathrm{G}_4 \rightarrow {}^3\mathrm{F}_2$            & $2800.0\,\text{s}^{-1}$                & $[2000, 4000]\,\text{s}^{-1}$   & Energy gap law ($p = 6.0$) \cite{layne1977, vandijk1983, jackson1998} \\
$K_{\text{CR0}}$           & Base cross-relaxation coefficient                                                  & $53580.0\,\text{s}^{-1}$               & $[35000, 75000]\,\text{s}^{-1}$ & F\"orster dipole coupling ($R_{\text{avg}} = 0.92\,\text{nm} < R_0$) \cite{dexter1953, auzel2004, moulton2009} \\
$K_{\text{CR1}}$           & ${}^1\mathrm{G}_4$ self-quenching rate                                             & $3270.0\,\text{s}^{-1}$                & $[1500, 5000]\,\text{s}^{-1}$   & Quenching via residual clusters ($\text{Al/Tm} \approx 3.0$) \cite{unger2025, peterka2011, leich2022oe, kamradek2019jphot} \\
{$K_{\mathrm{MPA}}$} & {Macroscopic multi-photon rate coefficient}           & {$4.71 \times 10^{-6}\,\text{mW}^{-3.3}\text{s}^{-1}$} & {$[3.5, 6.0] \times 10^{-6}$} & {Calibrated pumping rate ($m=3.3$) \cite{boyd2020, he2008, mangini2020}} \\
$\alpha_{\text{eff}}$      & Core thermal accumulation coefficient                                              & $9.31 \times 10^{-3}\,\text{K/mW}^2$  & $[6.5, 12.0] \times 10^{-3}$    & Lumped thermal parameter mapping local modes \cite{eaton2005, lancry2013} \\
$T_{\text{ph}}$            & Silica phonon temperature ($\hbar\omega = 1100\,\text{cm}^{-1}$)                   & $1584.0\,\text{K}$                     & --                              & Peak Si--O--Si stretching vibration \cite{layne1977, galeener1979} \\
$\beta_{\text{th}}$        & CR thermal assistance factor                                                       & $2.0$                                  & $[1.5, 2.8]$                    & Dexter-Grant phonon assistance factor \cite{grant1971, auzel2004} \\
$\eta_4$                   & NIR spectral overlap fraction                                                      & $0.28$                                 & $[0.22, 0.32]$                  & Convolution of   $804 \text{nm}$ band in $[787\text{--}797\,\text{nm}]$ \cite{unger2025} \\ 
\bottomrule
\end{tabularx}
\end{table*}

Numerical integration of Eqs.~(\ref{eq:ode1})--(\ref{eq:ode3}) using a stiff variable-order Radau IIA solver (Fig.~\ref{fig:timedyn}) indicates that populations reach steady-state equilibrium within $t < 2\,\text{ms}$, well within continuous experimental acquisition windows.

\begin{figure}[htbp]
  \centering
  \includegraphics[width=\columnwidth]{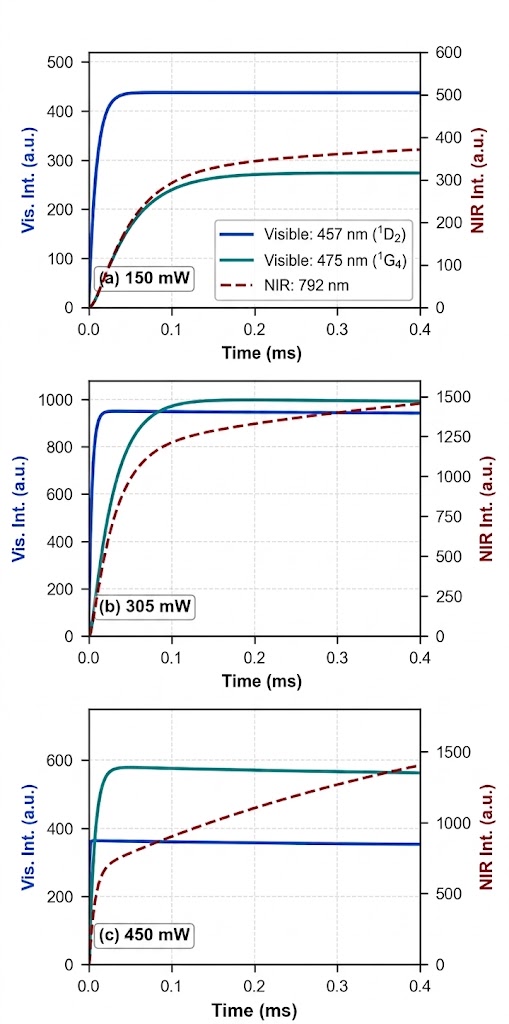}
  \caption{Calculated fractional population dynamics ($n_2, n_1, n_4$) across three regimes: (a) low power ($150\,\text{mW}$), (b) crossover ($280\,\text{mW}$), and (c) saturation ($450\,\text{mW}$). Populations reach equilibrium within 2~ms.}
  \label{fig:timedyn}
\end{figure}

\section{Results and discussion}

\subsection{Raw spectral characteristics and visible-NIR competition}
The acquired emission spectra as a function of launched average power are shown in Fig.~\ref{fig:spectra}. In the visible range (Fig.~\ref{fig:spectra}(a)), deep blue emission (${}^1\mathrm{D}_2 \rightarrow {}^3\mathrm{F}_4$, 457~nm) dominates below $240\,\text{mW}$. Above $260\,\text{mW}$, Cyan emission (${}^1\mathrm{G}_4 \rightarrow {}^3\mathrm{H}_6$, 475~nm) increases and then matches the 457 nm-line at a distinct crossover power threshold $P_{\text{cross}} \approx 283.1\,\text{mW}$. Beyond $320\,\text{mW}$, both blue channels saturate and undergo an asymmetric roll-off.

\begin{figure}[htbp]
  \centering
  \includegraphics[width=\columnwidth]{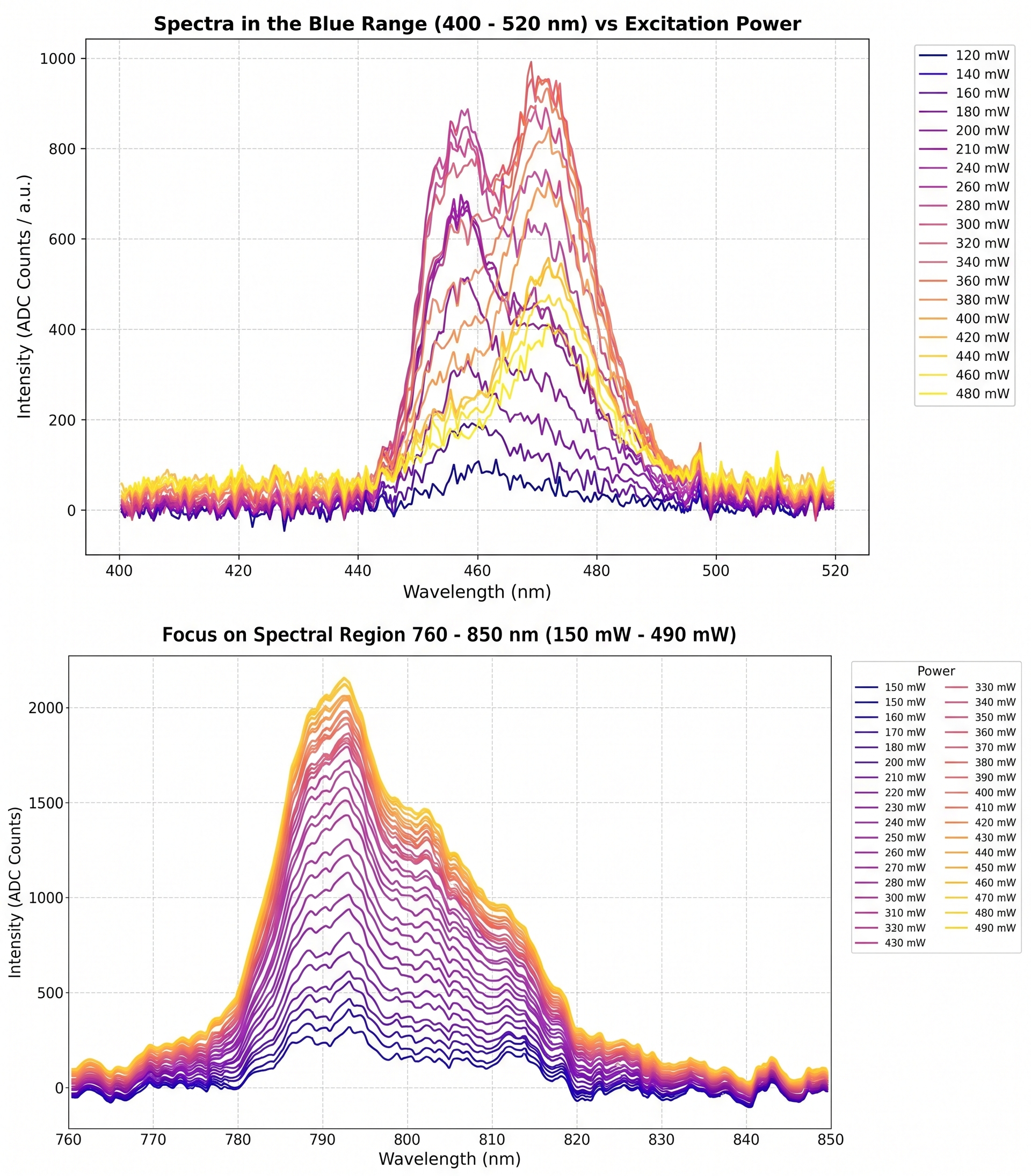}
  \caption{Power-dependent emission spectra acquired under ultrafast 1030-nm excitation: (a) Visible upconversion window showing the spectral crossover between deep blue (457~nm) and cyan (475~nm) lines; (b) Near-infrared window documenting the persistent, monotonic growth of the 792-nm band across the high-power visible roll-off regime.}
  \label{fig:spectra}
\end{figure}
In contrast to visible quenching, the near-infrared emission at $792\,\text{nm}$ (Fig.~\ref{fig:spectra}(b)) grows monotonically over the entire power range ($150\text{--}490\,\text{mW}$). 
\subsection{Low-power scaling analysis, power hysteresis, and reversible roll-off}
To examine the initial excitation order populating ${}^1\mathrm{D}_2$ prior to macroscopic saturation, a double-logarithmic regression was performed on 457-nm emission for $P \le 220\,\text{mW}$ (Fig.~\ref{fig:loglog_fit}). The experimental slope of $m = 3.37 \pm 0.22$ confirms a multi-photon excitation pathway populating ${}^1\mathrm{D}_2$ under spectral broadening conditions \cite{mangini2020, boyd2020}. Between $210\,\text{mW}$ and $240\,\text{mW}$, the deep blue intensity exhibits an intermediate plateau before continuing its climb toward maximum emission at $280\,\text{mW}$. This behavior reflects transient physical dynamics; indeed, internally generated blue luminescence transiently excites native precursor defect states (such as $\text{Al-OHC}$ and $E'$ centers) around 200--220~mW, introducing a prompt localized absorption.
Above 240~mW, cumulative thermal accumulation ($T_{\text{loc}} = T_0 + \alpha_{\text{eff}} P^2$) raises the core temperature, inducing dynamic thermal bleaching of these unstable color centers and restoring high transmission. Furthermore, this interval coincides with the onset of Self-Phase Modulation sidelobe restructuring, which temporarily redistributes on-resonance spectral power density before settling into asymptotic saturation \cite{agrawal2019}.

\begin{figure}[!htbp]
  \centering
  \includegraphics[width=\columnwidth]{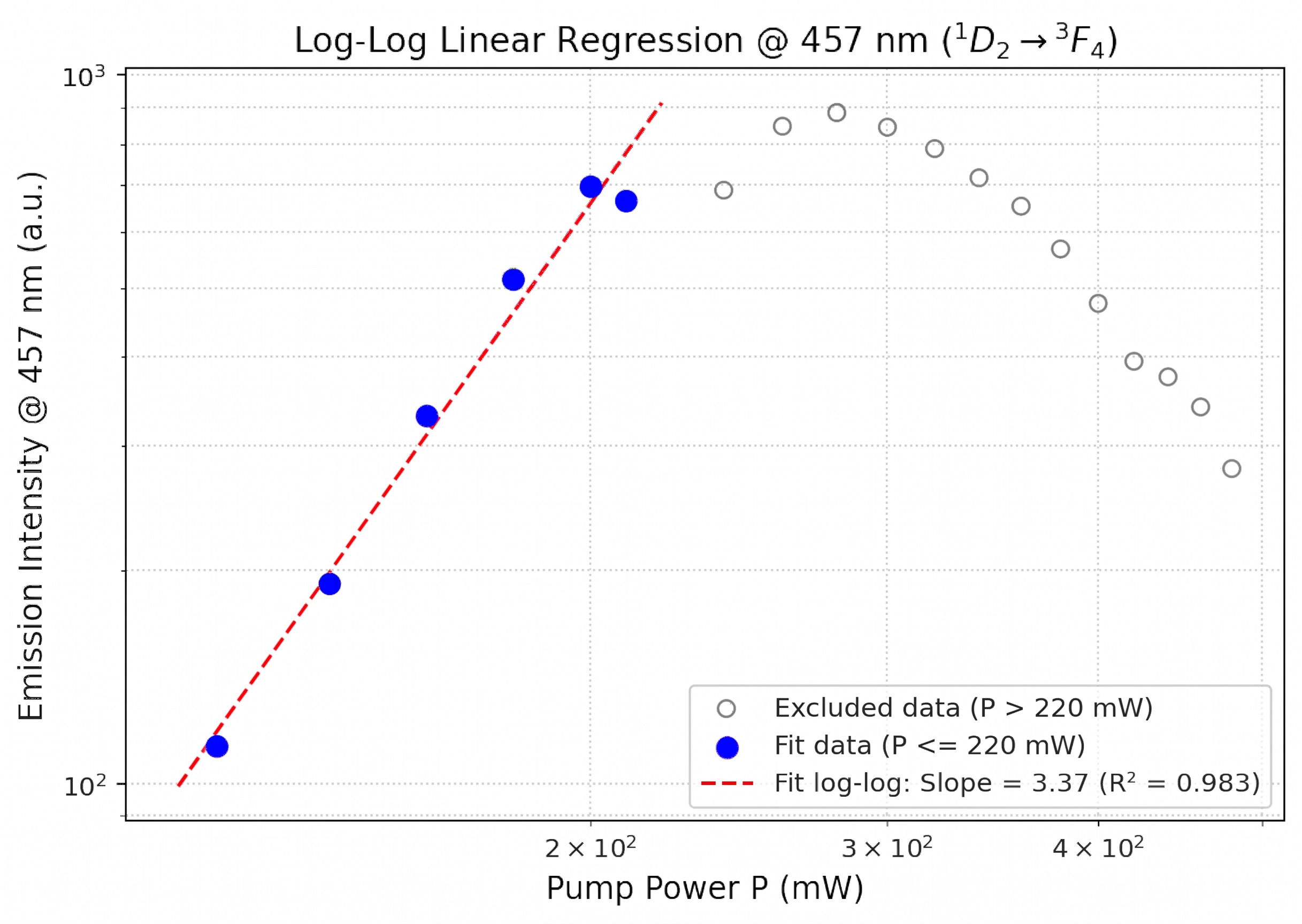}
  \caption{Double-logarithmic plot of Deep Blue emission (457~nm) versus launched power. Regression yields a slope of $m = 3.37 \pm 0.22$ ($R^2 = 0.983$) in the perturbative regime ($P \le 220\,\text{mW}$, solid circles), validating an SPM-broadened multi-photon process. Open circles indicate subsequent saturation, transient plateau, and high-power roll-off ($P > 220\,\text{mW}$).}
  \label{fig:loglog_fit}
\end{figure}

\subsection{Thermal phonon accumulation under repetition-rate variation}
To differentiate between thermal phonon accumulation and instantaneous peak optical intensity, the repetition rate was varied from $253\,\text{kHz}$ to $1008\,\text{kHz}$ at constant single-pulse energy ($E_p = \text{const}$, Fig.~\ref{fig:reprate}).

Under purely athermal excitation, emission intensity scales linearly with repetition rate ($\propto f_{\text{rep}}^1$) \cite{ferraro2021prj}. Instead, the observed signal respectively exhibits superlinear ($\propto f_{\text{rep}}^{1.5}$) and sublinear ($\propto f_{\text{rep}}^{0.6 - 0.9}$) dependencies in the NIR} and in the the blue ranges. Increasing $f_{\text{rep}}$ shortens pulse separation below thermal diffusion times, raising the core temperature and enhancing non-radiative transfer.

\begin{figure*}[!t]
  \centering
  \includegraphics[width=0.98\textwidth]{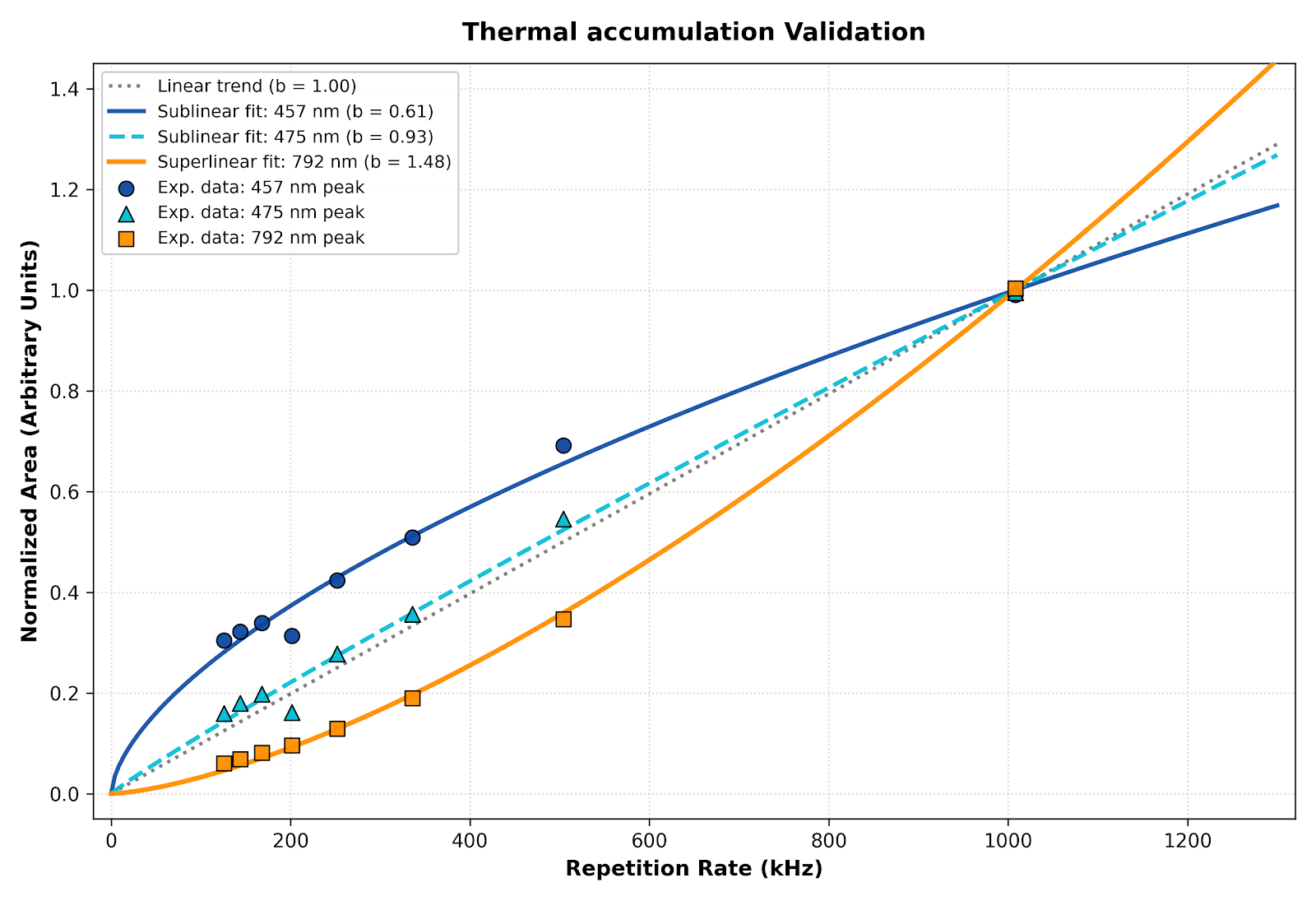}
  \caption{Integrated emission intensities as a function of laser repetition rate under constant pulse energy ($E_p = \text{const}$). Plotted scalings confirm cumulative thermal phonon accumulation. Data are fitted against the generic power law y=a$\cdot x^{b}$ }
  \label{fig:reprate}
\end{figure*}

\subsection{Simultaneous model validation across visible and NIR channels}

\begin{figure*}[!t]
  \centering
  \includegraphics[width=0.98\textwidth]{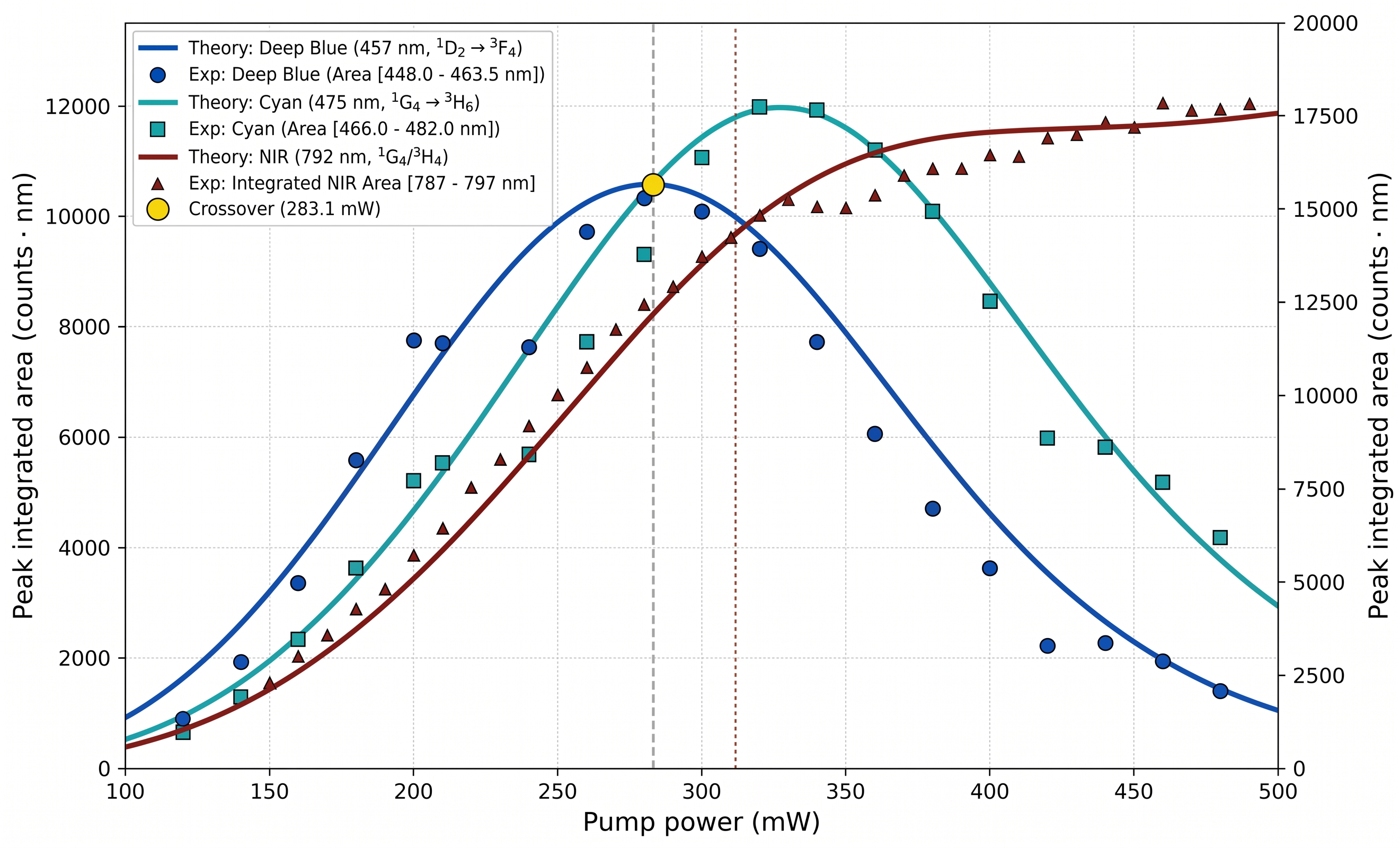}
  \caption{Simultaneous comparison of integrated experimental areas and rate-equation solutions: Deep Blue (457~nm, blue circles), Cyan (475~nm, cyan squares), and NIR (792~nm, brown triangles). The model accurately tracks the Deep Blue peak (280~mW), Cyan peak (320~mW), spectral crossover ($283.1\,\text{mW}$), and NIR bilinear slope transition. Emitted wavelengths maxima for each injected power were integrated over bands $\Delta\lambda \approx 15.0\,\text{nm}$.}
  \label{fig:comparison}
\end{figure*}

Figure~\ref{fig:comparison} compares integrated experimental intensities with rate-equation calculations across the entire pump range ($100\text{--}500\,\text{mW}$). Good quantitative agreement across all three spectral bands 
 is achieved;
\begin{enumerate}
  \item \textbf{Spectral crossover ($283.1\,\text{mW}$):} The model predicts the transition from ${}^1\mathrm{D}_2$ to ${}^1\mathrm{G}_4$ dominance, matching experimental data crossing between $280\,\text{mW}$ and $300\,\text{mW}$. The crossover is driven by non-radiative multi-phonon transfer ($W_{\text{MP2}} \propto [1+n_{\text{th}}]^{7.0}$) and thermally assisted cross-relaxation ($K_{\text{CR2}} \propto [1+2.0 n_{\text{th}}]$), transferring population to ${}^1\mathrm{G}_4$ as core temperature rises.
  \item \textbf{Thermal visible quenching:} Above 320~mW, multi-phonon drain from ${}^1\mathrm{G}_4$ ($W_{\text{MP1}} \propto [1+n_{\text{th}}]^{6.0}$) quenches both blue lines without empirical photodarkening terms.
  \item \textbf{NIR persistence:} Decays from ${}^1\mathrm{G}_4$ and cross-relaxation from ${}^1\mathrm{D}_2$ feed the metastable ${}^3\mathrm{H}_4$ level. Together with ground-state depletion, this reproduces the observed slope transition at $P_{\text{break}} \approx 312\,\text{mW}$.
\end{enumerate}

Noticeably, the experimental NIR data exhibit a minor local sagging (dip) between $330\,\text{mW}$ and $410\,\text{mW}$ before recovering the theoretical asymptotic trajectory above $420\,\text{mW}$. This subtle feature reflects a transient physical interplay: around $320\text{--}350\,\text{mW}$, the sharp multi-phonon quenching of ${}^1\mathrm{G}_4$ promptly terminates its direct contribution ($\gamma_{G4,\text{nir}}$), while population transiently bottlenecks in the long-lived terminal level ${}^3\mathrm{F}_4$ ($\tau \sim 500\,\mu\text{s}$), slightly depressing $n_0$ and the cross-relaxation feeding rates. Concurrently, thermal line broadening of the ${}^3\mathrm{H}_4 \rightarrow {}^3\mathrm{H}_6$ Stark manifold temporarily shifts a fraction of the emission tail outside the narrow $787\text{--}797\,\text{nm}$ bandpass window. Once $P > 420\,\text{mW}$, the ${}^1\mathrm{G}_4$ channel is fully extinguished and the multi-phonon cascade directly feeds ${}^3\mathrm{H}_4$, restoring the steady linear growth in full quantitative agreement with the rate-equation asymptote.

\section{Conclusions}

In summary, we have established a thermodynamically consistent rate-equation model coupled with guided-field analysis to elucidate the upconversion and near-infrared dynamics in ultrafast $\text{Tm}^{3+}$-doped aluminosilicate fibers under $1030\,\text{nm}$ excitation. Our results demonstrate that non-resonant multi-photon pumping into ${}^1\mathrm{D}_2$ is mediated by Kerr-induced beam confinement and SPM spectral broadening. Furthermore, we show that high-power visible quenching and spectral crossovers originate from thermal multi-phonon decay and cross-relaxation rather than permanent optical damage, while the ${}^3\mathrm{H}_4$ level maintains monotonic NIR growth. This work provides essential quantitative bounds for predicting thermal loads and optimizing power scaling in high-power thulium fiber devices.

\section*{Acknowledgment}
We acknowledge the support from the Italian Ministry of University and Research (MUR) through the Italian Science Fund (FIS2), the SOFT project (H53C24001530001), the European Union, and the Czech Republic state budget under the LasApp $CZ.02.01.01/00/22 008/0004573$ project.

\appendix

\section{Effective microscopic pumping rate and hysteresis sanity check}
\label{app:k3pa}
{Under ultrafast pulsed excitation, the multi-photon transition rate in the core scales with the effective guided core power $P_{\text{core}}$ as $R_{\text{MPA,core}} = K_{\text{core}} P_0 (P_{\text{core}}/P_0)^{3.3}$ ($P_0 = 1.0\,\text{mW}$), where evaluating the microscopic 3PA cross-section ($\sigma_{\text{3PA}} \approx 1.43 \times 10^{-82}\,\text{cm}^6\text{s}^2$) at the SPM-broadened resonance ($\lambda_{\text{res}} = 3 \times 355.0\,\text{nm} = 1065.0\,\text{nm}$) yields $K_{\text{core}} \approx 3.39 \times 10^{-5}\,\text{mW}^{-3.3}\,\text{s}^{-1}$. Accounting for core coupling $P_{\text{core}} = \eta_{\text{coup}} P_{\text{launch}}$, the macroscopic pumping rate referenced to launched power becomes}:
\begin{equation}
  {R_{\text{MPA}}(P_{\text{launch}}) = \left[ K_{\text{core}} \, \eta_{\text{coup}}^{3.3} \right] P_{\text{launch}}^{3.3} = K_{\mathrm{MPA}} P_{\text{launch}}^{3.3}}
\end{equation}
With $\eta_{\text{coup}} = 0.55$, the coefficient evaluates directly to:
\begin{equation}
  {K_{\mathrm{MPA}} = 3.39 \times 10^{-5} \times (0.55)^{3.3} \approx 4.71 \times 10^{-6}\,\text{mW}^{-3.3}\,\text{s}^{-1}}
\end{equation}
{confirming agreement with the calibrated rate-equation parameter.}

\begin{figure}[htbp]
  \centering
  \includegraphics[width=\columnwidth]{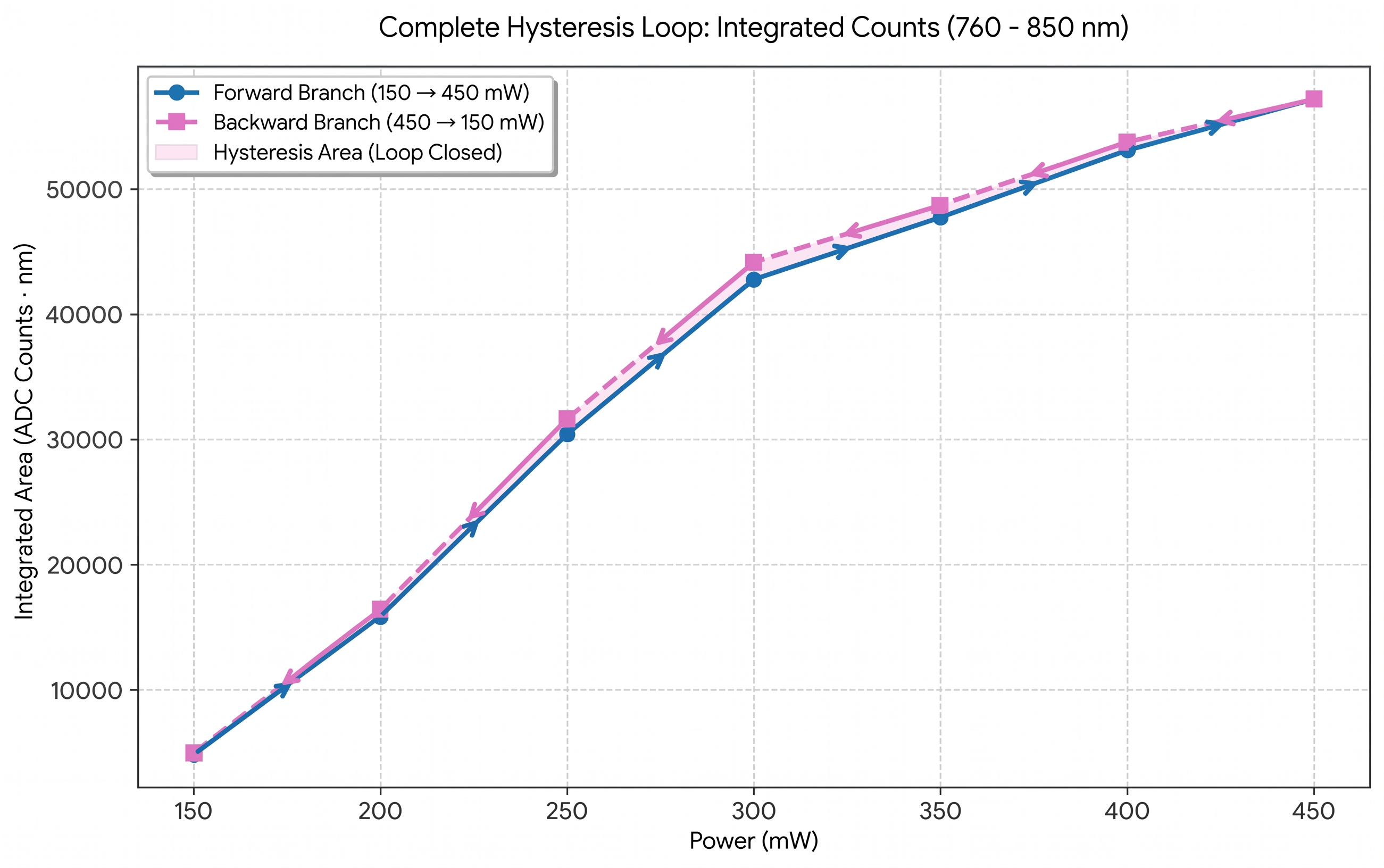}
  \caption{{Power hysteresis test performed by cycling launched power continuously upwards ($120 \rightarrow 480\,\text{mW}$) and downwards ($480 \rightarrow 120\,\text{mW}$ ). Because guided peak intensities remain bounded below breakdown, the overlapping sweeps confirm the absence of permanent structural solarization.}}
  \label{fig:hysteresis}
\end{figure}

{To make sure that the subsequent visible roll-off does not originate from permanent photochemical damage, cycling hysteresis sweeps were conducted by continuously ramping launched power upward ($120 \rightarrow 480\,\text{mW}$) and downward ($480 \rightarrow 120\,\text{mW}$), as shown in Fig.~\ref{fig:hysteresis}. The forward and backward trajectories overlap without measurable hysteresis, proving that high-power visible quenching and the near-infrared knee are governed entirely by reversible thermal equilibria (multi-phonon decay and cross-relaxation) rather than irreversible structural solarization.}

\section{Adiabatic elimination and rate-equation model terms}
\label{app:qssa}
The variables $n_i(t)$ represent dimensionless fractional populations normalized to total ion density $N_{\text{Tm}}$:
\begin{equation}
  n_i(t) = \frac{N_i(t)}{N_{\text{Tm}}}, \quad \sum_{i=0}^4 n_i(t) = 1
\end{equation}
associated with the physical energy manifolds:
\begin{itemize}
  \item $n_0$: Ground-state manifold ${}^3\mathrm{H}_6$.
  \item $n_1$: Cyan emitting level ${}^1\mathrm{G}_4$ ($475\,\text{nm}$).
  \item $n_2$: Deep Blue emitting level ${}^1\mathrm{D}_2$ ($457\,\text{nm}$).
  \item $n_3$: Metastable bottleneck state ${}^3\mathrm{F}_4$.
  \item $n_4$: Intermediate level ${}^3\mathrm{H}_4$ contributing to 792-nm emission.
\end{itemize}

Intermediate levels ${}^3\mathrm{F}_{2,3}$ and ${}^3\mathrm{H}_5$ lie closely above ${}^3\mathrm{H}_4$ ($\Delta E \approx 1980\,\text{cm}^{-1}$) and ${}^3\mathrm{F}_4$ ($\Delta E \approx 2220\,\text{cm}^{-1}$). Non-radiative decays across these small gaps exceed $10^9\,\text{s}^{-1}$ ($\tau < 0.5\,\text{ns}$) \cite{layne1977, unger2025}. Applying the Quasi-Steady-State Approximation:
\begin{equation}
  \frac{dn({}^3\mathrm{F}_{2,3})}{dt} \approx 0, \quad \frac{dn({}^3\mathrm{H}_5)}{dt} \approx 0
\end{equation}
Transitions feeding ${}^3\mathrm{F}_{2,3}$ cascade instantaneously into ${}^3\mathrm{H}_4$, while transitions feeding ${}^3\mathrm{H}_5$ cascade directly into ${}^3\mathrm{F}_4$.

The microscopic rate coefficients governing the dynamical system are defined as follows:
\begin{itemize}
  \item \textbf{Transition ${}^1\mathrm{D}_2 \rightarrow {}^1\mathrm{G}_4$ ($W_{\text{MP2}}$):} Governed by the modified energy gap law across $\Delta E \approx 6709\,\text{cm}^{-1}$ \cite{unger2025}: $W_{\text{MP2}} = W_{0,2} [1 + n_{\text{th}}]^{7.0}$ with base rate $W_{0,2} = 23180\,\text{s}^{-1}$, where the effective phonon order $p = 7.0$ stems from the altered local vibrational cutoff ($\hbar\omega_{\text{eff}} \approx 960\,\text{cm}^{-1}$) of $[\text{AlO}_4]^-$ units \cite{arai1986, peterka2011, leich2022oe, kamradek2019jphot}.
    
  \item \textbf{Multi-phonon drain ${}^1\mathrm{G}_4 \rightarrow {}^3\mathrm{F}_2$ ($W_{\text{MP1}}$):} Governed by $W_{\text{MP1}} = W_{0,1} [1 + n_{\text{th}}]^{6.0}$ across $\Delta E \approx 6080\,\text{cm}^{-1}$ with $W_{0,1} = 2800\,\text{s}^{-1}$ \cite{jackson1998, simpson2006}.
\end{itemize}

For an average doping concentration of $13872\,\text{ppm}$ atomic ($N_{\text{Tm}} \approx 3.05 \times 10^{20}\,\text{ions/cm}^3$), the mean inter-ionic spacing is $R_{\text{avg}} = (3 / 4\pi N_{\text{Tm}})^{1/3} \approx 0.92\,\text{nm}$. Comparing with the critical F\"orster radius $R_0 = 1.08 \pm 0.05\,\text{nm} \approx 1.1\,\text{nm}$ \cite{dexter1953, auzel2004, unger2025, agger2006}, the transfer efficiency exceeds $P_{\text{ET}} = [1 + (R_{\text{avg}}/R_0)^6]^{-1} > 80\%$, confirming the strong dipole-coupling regime \cite{jackson1998, moulton2009}:
\begin{itemize}
  \item \textbf{Base cross-relaxation rate ($K_{\text{CR0}}$):} Scales as $K_{\text{CR}} \approx \tau_0^{-1} (R_0 / R_{\text{avg}})^6 \approx (20000\,\text{s}^{-1}) \times (1.1 / 0.92)^6 \approx 5.8 \times 10^4\,\text{s}^{-1}$, validating the base coefficient $K_{\text{CR0}} = 53580\,\text{s}^{-1}$ for ${}^1\mathrm{D}_2 + {}^3\mathrm{H}_6 \rightarrow {}^1\mathrm{G}_4 + {}^3\mathrm{H}_4$ \cite{jackson1998, moulton2009}.
    
  \item \textbf{Phonon assistance parameter ($\beta_{\text{th}}$):} Compensates for the energy deficit:
  \begin{equation}
    \Delta E_{\text{mismatch}} = [E({}^1\mathrm{D}_2) - E({}^1\mathrm{G}_4)] - [E({}^3\mathrm{H}_4) - E({}^3\mathrm{H}_6)] \approx 759\,\text{cm}^{-1}
  \end{equation}
  Under Dexter-Grant theory \cite{dexter1953, grant1971}, the temperature-assisted rate is:
  \begin{equation}
    K_{\text{CR2}}(T) = K_{\text{CR0}} \left[ 1 + \beta_{\text{th}} n_{\text{th}}(T_{\text{loc}}) \right]
  \end{equation}
  with $\beta_{\text{th}} = 2.0$, reflecting matrix phonon absorption in the $700\text{--}800\,\text{cm}^{-1}$ acoustic/optical band \cite{grant1971, auzel2004}.
    
  \item \textbf{Concentration self-quenching ($K_{\text{CR1}}$):} The secondary rate $K_{\text{CR1}} = 3270\,\text{s}^{-1}$ models concentration self-quenching from ${}^1\mathrm{G}_4$ (${}^1\mathrm{G}_4 + {}^3\mathrm{H}_6 \rightarrow {}^3\mathrm{H}_4 + {}^3\mathrm{F}_4$) induced by residual paired or clustered ions ($5\text{--}8\%$) at $\text{Al/Tm} \approx 3.0$ \cite{arai1986, leich2022oe, unger2025, kamradek2019jphot}.
\end{itemize}

\section{Thermal model, heat accumulation, and local phonon coupling}
\label{app:thermal}
Because the inter-pulse separation ($t_{\text{inter}} \approx 0.99\,\mu\text{s}$) is an order of magnitude shorter than the core thermal diffusion time:
\begin{equation}
  \tau_{\text{diff}} = \frac{w_0^2}{4 D_{\text{th}}} \approx \frac{(6.335 \times 10^{-6}\,\text{m})^2}{4 \times 8.4 \times 10^{-7}\,\text{m}^2/\text{s}} \approx 11.9\,\mu\text{s}
\end{equation}
deposited thermal energy cannot escape between consecutive pulses, establishing cumulative steady-state heat accumulation \cite{eaton2005}.

Under steady radial heat conduction across a cylindrical silica cladding of radius $r_{\text{clad}} = 64\,\mu\text{m}$ and interaction length $L$:
\begin{equation}
  \frac{1}{r} \frac{d}{dr} \left( r \frac{dT}{dr} \right) = 0, \quad w_0 \le r \le r_{\text{clad}}
\end{equation}
Integrating from the core mode radius $w_0$ to the outer boundary $r_{\text{clad}}$ yields the core-to-surface temperature rise:
\begin{equation}
  \Delta T = T_{\text{core}} - T_{\text{surface}} = \frac{Q_{\text{heat}}}{2\pi k_{\text{th}} L} \ln\left( \frac{r_{\text{clad}}}{w_0} \right)
\end{equation}
where $k_{\text{th}} = 1.38\,\text{W/(m}\cdot\text{K)}$ is the thermal conductivity of fused silica.

Because linear background absorption at $1030\,\text{nm}$ is negligible ($\alpha_0 < 0.01\,\text{dB/m}$), internal dissipation $Q_{\text{heat}}$ is dominated by non-radiative relaxation following multi-photon excitation:
\begin{equation}
  Q_{\text{heat}} \propto I_{\text{peak}} P_{\text{avg}} \propto \frac{P_{\text{avg}}^2}{f_{\text{rep}}\tau_{\text{FWHM}}A_{\text{eff}}} \propto P_{\text{launch}}^2
\end{equation}
yielding the quadratic functional temperature law:
\begin{equation}
  T_{\text{loc}}(P) = T_0 + \alpha_{\text{eff}} P_{\text{launch}}^2
\end{equation}
with $T_0 = 293.15\,\text{K}$. 

Operationally, $T_{\text{loc}}$ represents an effective local vibrational coordinate rather than a macroscopic bulk temperature. {The occupation number of the vibrational field follows the Bose--Einstein distribution}:
\begin{equation}
  {n_{\text{th}}(T_{\text{loc}}) = \frac{1}{\exp\left(\frac{T_{\text{ph}}}{T_{\text{loc}}}\right) - 1}}
\end{equation}
While $n_{\text{th}}$ is conventionally formulated with respect to the high-energy asymmetric stretching vibration of fused silica ($T_{\text{ph}} = 1584\,\text{K}$, $\hbar\omega_{\text{bulk}} \approx 1100\,\text{cm}^{-1}$), non-radiative multi-phonon bridging in aluminosilicate fibers is strongly mediated by localized low-frequency modes of $[\text{AlO}_4]^-$ coordination tetrahedra ($350\text{--}500\,\text{cm}^{-1}$, $T_{\text{ph,local}} \approx 500\text{--}720\,\text{K}$) directly surrounding the $\text{Tm}^{3+}$ solvation shell \cite{arai1986, leich2022oe, peterka2011}. Within our lumped model, the effective coefficient $\alpha_{\text{eff}} = 9.31 \times 10^{-3}\,\text{K/mW}^2$ maps the localized thermal occupancy onto these low-frequency coordination modes. Its exact numerical value is constrained by the kinematic maximum of the 457-nm emission:
\begin{equation}
  \left. \frac{dn_2}{dP} \right|_{P = 280\,\text{mW}} = 0
\end{equation}
which balances cubic optical pumping against temperature-accelerated multi-phonon decay.

\bibliographystyle{elsarticle-num}
\bibliography{references}

\begin{thebibliography}{10}
\expandafter\ifx\csname url\endcsname\relax
  \def\url#1{\texttt{#1}}\fi
\expandafter\ifx\csname urlprefix\endcsname\relax\def\urlprefix{URL }\fi
\expandafter\ifx\csname href\endcsname\relax
  \def\href#1#2{#2} \def\path#1{#1}\fi

\bibitem{digo2001}
M.~J.~F. Digonnet (Ed.), Rare-Earth-Doped Fiber Lasers and Amplifiers, 2nd Edition, Marcel Dekker, New York, 2001.

\bibitem{khamis2016}
M.~A. Khamis, K.~Ennser, Theoretical model of a thulium-doped fiber amplifier pumped at 1570 nm and 793 nm in the presence of cross relaxation, Journal of Lightwave Technology 34~(24) (2016) 5675--5681.
\newblock \href {https://doi.org/10.1109/JLT.2016.2631635} {\path{doi:10.1109/JLT.2016.2631635}}.

\bibitem{duclos1995}
F.~Duclos, P.~Urquhart, Thulium-doped {ZBLAN} blue upconversion fiber laser: theory, Journal of the Optical Society of America B 12~(4) (1995) 709--717.
\newblock \href {https://doi.org/10.1364/JOSAB.12.000709} {\path{doi:10.1364/JOSAB.12.000709}}.

\bibitem{varak2024}
P.~Va{\v{r}}{\'a}k, et~al., Fluorescence lifetime of highly rare-earth-doped silica optical fibers -- the influence of composition and fabrication processing, in: 2024 24th International Conference on Transparent Optical Networks (ICTON), 2024, pp. 1--4.
\newblock \href {https://doi.org/10.1109/icton62926.2024.10647713} {\path{doi:10.1109/icton62926.2024.10647713}}.

\bibitem{gaida2017}
C.~Gaida, M.~Gebhardt, F.~Stutzki, C.~Jauregui, J.~Limpert, A.~T{\"u}nnermann, 90 fs pulses with $>$5 {GW} peak power from a high repetition rate {Tm}-doped fiber {CPA} system, in: Advanced Solid State Lasers, 2017, p. ATh3A.5.
\newblock \href {https://doi.org/10.1364/ASSL.2017.ATH3A.5} {\path{doi:10.1364/ASSL.2017.ATH3A.5}}.

\bibitem{yinzi2020}
L.~Yinzi, et~al., Thermal bleaching of photodarkening and heat-induced loss and spectral broadening in {Tm}$^{3+}$-doped fibers, Optics Express 28~(15) (2020) 21845--21853.
\newblock \href {https://doi.org/10.1364/OE.397356} {\path{doi:10.1364/OE.397356}}.

\bibitem{tao2013}
M.~Tao, Q.~Huang, T.~Yu, P.~Yang, W.~Chen, X.~Ye, Cross relaxation in {Tm}-doped fiber lasers, in: Selected Papers of the Photoelectronic Technology Committee Conferences held November 2012 and June 2013, Vol. 8796 of Proceedings of SPIE, 2013, p. 87961A.
\newblock \href {https://doi.org/10.1117/12.2011078} {\path{doi:10.1117/12.2011078}}.

\bibitem{moulton2009}
P.~F. Moulton, G.~A. Rines, E.~V. Slobodtchikov, K.~F. Wall, G.~Frith, A.~L. Carter, B.~Samson, {Tm}-doped fiber lasers: Fundamentals to high power, IEEE Journal of Selected Topics in Quantum Electronics 15~(1) (2009) 85--92.
\newblock \href {https://doi.org/10.1109/JSTQE.2008.2010719} {\path{doi:10.1109/JSTQE.2008.2010719}}.

\bibitem{Paschotta__quantum_efficiency}
R.~Paschotta, Quantum efficiency (2008).
\newblock \href {https://doi.org/10.61835/fxl} {\path{doi:10.61835/fxl}}.

\bibitem{simpson2008}
D.~A. Simpson, G.~W. Baxter, S.~F. Collins, W.~E.~K. Gibbs, P.~W. France, L.~J. Baskiotis, Visible and near infra-red up-conversion in {Tm}$^{3+}$/{Yb}$^{3+}$ co-doped silica fibres under 980 nm excitation, Optics Express 16~(18) (2008) 13781--13799.
\newblock \href {https://doi.org/10.1364/OE.16.013781} {\path{doi:10.1364/OE.16.013781}}.

\bibitem{tanabe1998}
S.~Tanabe, E.~Snitzer, B.~Cole, Blue upconversion characteristics of thulium-doped silica fiber with high germania content, Japanese Journal of Applied Physics 37~(Suppl. 37-1) (1998) 81--83.
\newblock \href {https://doi.org/10.7567/JJAPS.37S1.81} {\path{doi:10.7567/JJAPS.37S1.81}}.

\bibitem{auzel2004}
F.~Auzel, Upconversion and anti-stokes processes with f and d ions in solids, Chemical Reviews 104~(1) (2004) 139--174.
\newblock \href {https://doi.org/10.1021/cr020357g} {\path{doi:10.1021/cr020357g}}.

\bibitem{barber1994}
P.~R. Barber, C.~J. Mackechnie, R.~D.~T. Lauder, H.~M. Pask, A.~C. Tropper, D.~C. Hanna, S.~D. Butterworth, M.~J. McCarthy, J.~L. Archambault, S.~T. Davey, All solid-state blue room-temperature thulium-doped upconversion fibre laser, in: Compact Blue-Green Lasers, 1994.
\newblock \href {https://doi.org/https://doi.org/10.1364/CBGL.1994.CFA.3} {\path{doi:https://doi.org/10.1364/CBGL.1994.CFA.3}}.

\bibitem{zhao2010}
P.~Zhao, J.~Liu, C.~Zhao, H.~Yang, J.~Wen, The slope efficiency of 2$\mu$m thulium doped fiber laser, in: High-Power Lasers and Applications V, Vol. 7843 of Proceedings of SPIE, 2010, p. 784306.
\newblock \href {https://doi.org/10.1117/12.868620} {\path{doi:10.1117/12.868620}}.

\bibitem{frith2010}
G.~Frith, A.~Carter, B.~Samson, J.~Faroni, K.~Tankala, G.~Oulundsen, D.~Shenk, Mitigation of photodegradation in 790nm-pumped {Tm}-doped fibers, in: Fiber Lasers VII: Technology, Systems, and Applications, Vol. 7580 of Proceedings of SPIE, 2010, p. 75800A.
\newblock \href {https://doi.org/10.1117/12.846230} {\path{doi:10.1117/12.846230}}.

\bibitem{pal2013}
A.~Pal, Thulium-doped fibre laser in the 2 $\mu$m wavelength region for gas sensing, Ph.D. thesis, University of Southampton (2013).

\bibitem{jlumin2023}
R.~Filosa, et~al., Nonlinear optical effects in natural topaz, Journal of Luminescence 261 (2023) 120076.
\newblock \href {https://doi.org/10.1016/j.jlumin.2023.120076} {\path{doi:10.1016/j.jlumin.2023.120076}}.

\bibitem{yang2018}
Y.~Yang, et~al., Blue upconversion in {Yb}$^{3+}$/{Tm}$^{3+}$ co-doped silica fiber based on glass phase-separation technology, Applied Physics A 124~(2) (2018) 205.
\newblock \href {https://doi.org/10.1007/S00339-018-1638-X} {\path{doi:10.1007/S00339-018-1638-X}}.

\bibitem{xia2015}
C.~Xia, G.~Zhou, J.~Liu, W.~Zhang, Y.~Han, J.~Yuan, Blue and near-infrared up-conversion in double-air-cladding {Tm}$^{3+}$-doped silica fiber under 1064 nm excitation, Optical and Quantum Electronics 47~(11) (2015) 3435--3444.
\newblock \href {https://doi.org/10.1007/S11082-015-0219-X} {\path{doi:10.1007/S11082-015-0219-X}}.

\bibitem{buckthorpe2025}
M.~Buckthorpe, S.~D. Jackson, Revisiting energy transfer in {Tm}-doped aluminosilicate glass fiber {I}: introducing a new model, Optical Materials Express 15~(10) (2025) 2576--2592.
\newblock \href {https://doi.org/10.1364/OME.573889} {\path{doi:10.1364/OME.573889}}.

\bibitem{kamradek2021}
M.~Kamr{\'a}dek, et~al., Energy transfer coefficients in thulium-doped silica fibers, Optical Materials Express 11~(6) (2021) 1805--1814.
\newblock \href {https://doi.org/10.1364/OME.427456} {\path{doi:10.1364/OME.427456}}.

\bibitem{unger2025}
S.~Unger, M.~Leich, A.~Schwuchow, M.~Lorenz, R.~M{\"u}ller, A.~Pratiwi, J.~Kobelke, A.~Lorenz, J.~Dellith, C.~Kr{\"a}nkel, M.~J{\"a}ger, Absorption and emission properties of thulium doped glasses in optical fibers, Optical Materials Express 15~(3) (2025) 446--464.
\newblock \href {https://doi.org/10.1364/OME.547199} {\path{doi:10.1364/OME.547199}}.

\bibitem{messias2002}
D.~Messias, M.~Vermelho, M.~de~Araujo, A.~Gouveia-Neto, J.~Aitchison, Blue energy upconversion emission in thulium-doped-sio/sub 2/-p/sub 2/o/sub 5/ channel waveguides excited at 1.064 /spl mu/m, IEEE Journal of Quantum Electronics 38~(12) (2002) 1647--1650.
\newblock \href {https://doi.org/10.1109/JQE.2002.805108} {\path{doi:10.1109/JQE.2002.805108}}.

\bibitem{ferraro2021prj}
M.~Ferraro, F.~Mangini, A.~D. Luca, M.~Zitelli, A.~Tonello, V.~Couderc, S.~Wabnitz, Femtosecond nonlinear losses in multimode optical fibers, Photonics Research 9~(12) (2021) 2443--2453.
\newblock \href {https://doi.org/10.1364/PRJ.425878} {\path{doi:10.1364/PRJ.425878}}.

\bibitem{mangini2020}
F.~Mangini, et~al., Multiphoton-absorption-driven up-conversion in silica optical fibers under ultrafast infrared excitation, Physical Review Applied 14~(5) (2020) 054063.
\newblock \href {https://doi.org/10.1103/PhysRevApplied.14.054063} {\path{doi:10.1103/PhysRevApplied.14.054063}}.

\bibitem{ferraro2024}
M.Ferraro, et~al., Observation of visible upconversion luminescence of soft glass multimode fibers, Fibers 2~(15) (2024).
\newblock \href {https://doi.org/10.3390/fib12020015} {\path{doi:10.3390/fib12020015}}.

\bibitem{ferraro2025}
M.~Ferraro, et~al., Extreme nonlinear optics in optical fibers (2025).
\newblock \href {https://doi.org/https://arxiv.org/abs/2512.25046} {\path{doi:https://arxiv.org/abs/2512.25046}}.

\bibitem{kamradek2019jphot}
M.~Kamr{\'a}dek, I.~Ka{\v{s}}{\'i}k, J.~Aubrecht, J.~Mr{\'a}zek, O.~Podrazk{\'y}, J.~Cajzl, P.~Va{\v{r}}{\'a}k, V.~Kube{\v{c}}ek, P.~Peterka, P.~Honz{\'a}tko, Nanoparticle and solution doping for efficient holmium fiber lasers, IEEE Photonics Journal 11~(5) (2019) 7103610.
\newblock \href {https://doi.org/10.1109/JPHOT.2019.2940747} {\path{doi:10.1109/JPHOT.2019.2940747}}.

\bibitem{leich2022oe}
M.~Leich, et~al., Thermal stability and enhanced spectroscopic response of highly thulium-doped aluminosilicate optical fibers, Optics Express 30~(6) (2022) 9123--9138.
\newblock \href {https://doi.org/10.1364/OE.449643} {\path{doi:10.1364/OE.449643}}.

\bibitem{peterka2011}
P.~Peterka, et~al., Theoretical modeling of fiber laser at 810 nm based on thulium-doped silica fibers with enhanced ${}^3\text{H}_4$ level lifetime, Optics Express 19~(3) (2011) 2773--2781.
\newblock \href {https://doi.org/10.1364/OE.19.002773} {\path{doi:10.1364/OE.19.002773}}.

\bibitem{eaton2005}
S.~M. Eaton, et~al., Heat accumulation effects in femtosecond laser-written waveguides with variable repetition rate, Optics Express 13~(12) (2005) 4708--4716.
\newblock \href {https://doi.org/10.1364/OPEX.13.004708} {\path{doi:10.1364/OPEX.13.004708}}.

\bibitem{lancry2013}
M.~Lancry, B.~Poumellec, Uv laser processing and multiphoton densification of silica glasses, Physics Reports 523~(4) (2013) 207--229.
\newblock \href {https://doi.org/10.1016/j.physrep.2012.09.008} {\path{doi:10.1016/j.physrep.2012.09.008}}.

\bibitem{elagmy2014}
R.~M. El-Agmy, N.~Al-Hosiny, S.~Abdallah, M.~S. Abdel-Aal, Generation of short wavelength in up-conversion of {Tm}$^{+3}$ doped fluoride glass and its application in fiber lasers, Journal of Modern Physics 5~(3) (2014) 123--127.
\newblock \href {https://doi.org/10.4236/JMP.2014.53021} {\path{doi:10.4236/JMP.2014.53021}}.

\bibitem{broer1993}
M.~M. Broer, D.~M. Krol, D.~J. DiGiovanni, Highly nonlinear, photoinduced absorption in {Tm}$^{3+}$-doped aluminosilicate silica fiber, Optics Letters 18~(10) (1993) 799--801.
\newblock \href {https://doi.org/10.1364/OL.18.000799} {\path{doi:10.1364/OL.18.000799}}.

\bibitem{broer1993ol}
M.~M. Broer, et~al., Photoinduced absorption dynamics in rare-earth-doped aluminosilicate fibers, Optics Letters 18~(24) (1993) 2105--2107.
\newblock \href {https://doi.org/10.1364/OL.18.002105} {\path{doi:10.1364/OL.18.002105}}.

\bibitem{lupi2014}
J.-F. Lupi, M.~Vermillac, W.~Blanc, B.~Dussardier, Spectroscopie du thulium pomp{\'e} {\`a} 1070 nm dans une fibre alumino-silicate. {F}luorescence, photo-noircissement et simulation, in: 33{\`e}mes Journ{\'e}es Nationales d'Optique Guid{\'e}e (JNOG), 2014.

\bibitem{fu2010}
Y.~Fu, J.~Chen, Theoretical modelling of dual-wavelength pumped {Yb}$^{3+}$--{Tm}$^{3+}$ co-doped silica fibre laser, Journal of Optics 12~(9) (2010) 095707.
\newblock \href {https://doi.org/10.1088/2040-8978/12/9/095707} {\path{doi:10.1088/2040-8978/12/9/095707}}.

\bibitem{tse2015}
C.~H. Tse, All fibre laser source and specialty fibre for 2$\mu$m laser applications, Ph.D. thesis, Nanyang Technological University (2015).
\newblock \href {https://doi.org/10.32657/10356/64883} {\path{doi:10.32657/10356/64883}}.

\bibitem{simpson2006}
D.~A. Simpson, et~al., Energy transfer up-conversion in {Tm}$^{3+}$-doped silica fiber, Journal of Non-Crystalline Solids 352~(2) (2006) 136--141.
\newblock \href {https://doi.org/10.1016/J.JNONCRYSOL.2005.11.019} {\path{doi:10.1016/J.JNONCRYSOL.2005.11.019}}.

\bibitem{jackson1998}
S.~D. Jackson, T.~A. King, Theoretical modeling of {Tm}-doped silica fiber lasers, IEEE Journal of Quantum Electronics 34~(9) (1998) 1578--1587.
\newblock \href {https://doi.org/10.1109/50.762916} {\path{doi:10.1109/50.762916}}.

\bibitem{pal2010}
A.~Pal, et~al., Energy-transfer parameters in a {Tm}/{Yb} doped single mode silica fiber, Journal of the Optical Society of America B 27~(12) (2010) 2714--2720.
\newblock \href {https://doi.org/10.1364/JOSAB.27.002714} {\path{doi:10.1364/JOSAB.27.002714}}.

\bibitem{lupi2016}
J.-F. Lupi, et~al., Steady photo-darkening of thulium alumino-silicate fibers pumped at 1.07 $\mu$m: Quantitative effect of lanthanum, cerium and thulium, Optics Letters 41~(12) (2016) 2771--2774.
\newblock \href {https://doi.org/10.1364/OL.41.002771} {\path{doi:10.1364/OL.41.002771}}.

\bibitem{arai1986}
K.~Arai, H.~Namikawa, K.~Kumata, T.~Honda, Y.~Ishii, T.~Handa, Aluminum or phosphorus co-doping effects on the fluorescence and structural properties of neodymium-doped silica glass, Journal of Applied Physics 59~(10) (1986) 3430--3436.
\newblock \href {https://doi.org/10.1063/1.336810} {\path{doi:10.1063/1.336810}}.

\bibitem{boyd2020}
R.~W. Boyd, Nonlinear Optics, 4th Edition, Academic Press, London, 2020.

\bibitem{he2008}
G.~S. He, L.~S. Tan, Q.~Zheng, P.~N. Prasad, Multiphoton absorbing materials: Molecular designs, characterizations, and applications, Chemical Reviews 108~(4) (2008) 1245--1330.
\newblock \href {https://doi.org/10.1021/cr050054x} {\path{doi:10.1021/cr050054x}}.

\bibitem{judd1962}
B.~R. Judd, Optical absorption intensities of rare-earth ions, Physical Review 127~(3) (1962) 750--761.
\newblock \href {https://doi.org/10.1103/PhysRev.127.750} {\path{doi:10.1103/PhysRev.127.750}}.

\bibitem{ofelt1962}
G.~S. Ofelt, Intensities of crystal spectra of rare-earth ions, The Journal of Chemical Physics 37~(3) (1962) 511--520.
\newblock \href {https://doi.org/10.1063/1.1701366} {\path{doi:10.1063/1.1701366}}.

\bibitem{caird1975}
J.~A. Caird, L.~G. DeShazer, J.~Nella, Characteristics of room-temperature 2.3-$\mu$m laser emission from {Tm}$^{3+}$ in crystals and glasses, IEEE Journal of Quantum Electronics 11~(11) (1975) 874--881.
\newblock \href {https://doi.org/10.1109/JQE.1975.1068541} {\path{doi:10.1109/JQE.1975.1068541}}.

\bibitem{layne1977}
C.~B. Layne, W.~H. Lowdermilk, M.~J. Weber, Multiphonon relaxation of rare-earth ions in oxide glasses, Physical Review B 16~(1) (1977) 10--20.
\newblock \href {https://doi.org/10.1103/PhysRevB.16.10} {\path{doi:10.1103/PhysRevB.16.10}}.

\bibitem{vandijk1983}
J.~M.~F. van Dijk, M.~F.~H. Schuurmans, On the nonradiative and radiative decay rates and a modified energy gap law for 4f-4f transitions in rare-earth ions, The Journal of Chemical Physics 78~(9) (1983) 5317--5323.
\newblock \href {https://doi.org/10.1063/1.445485} {\path{doi:10.1063/1.445485}}.

\bibitem{dexter1953}
D.~L. Dexter, A theory of sensitized luminescence in solids, The Journal of Chemical Physics 21~(5) (1953) 836--850.
\newblock \href {https://doi.org/10.1063/1.1699044} {\path{doi:10.1063/1.1699044}}.

\bibitem{galeener1979}
F.~L. Galeener, Band limits and the vibrational spectra of tetrahedral glasses, Physical Review B 19~(8) (1979) 4292--4298.
\newblock \href {https://doi.org/10.1103/PhysRevB.19.4292} {\path{doi:10.1103/PhysRevB.19.4292}}.

\bibitem{grant1971}
W.~J.~C. Grant, Role of exchange interactions in energy transfer in rare-earth systems, Physical Review B 4~(2) (1971) 648--663.
\newblock \href {https://doi.org/10.1103/PhysRevB.4.648} {\path{doi:10.1103/PhysRevB.4.648}}.

\bibitem{agrawal2019}
G.~P. Agrawal, Nonlinear Fiber Optics, 6th Edition, Academic Press, San Diego, 2019.

\bibitem{agger2006}
S.~Agger, J.~H. Povlsen, Emission and absorption cross section of thulium doped silica fibers, Optics Express 14~(1) (2006) 50--57.
\newblock \href {https://doi.org/10.1364/OPEX.14.000050} {\path{doi:10.1364/OPEX.14.000050}}.

\end{thebibliography}

\end{document}